\documentstyle[preprint,fleqn,epsf,aps,prb]{revtex}
\tightenlines

\begin{document}

\draft
\preprint{}

\title{Electronic properties of ternary quasicrystals in one dimension}

\author{Nobuhisa Fujita and Komajiro Niizeki}

\address{Department of Physics, Graduate School of Science,\\
         Tohoku University, Sendai 980-8578, Japan}

\date{\today}

\maketitle

\begin{abstract}
The one-electron properties of a certain class of one-dimensional ternary quasicrystals are investigated. In particular, we show in detail the presence of a special kind of critical states called marginal critical states in these QCs. By the use of a real-space renormalization-group method, it is shown that the scaling properties of marginal critical states are characterized by stretched exponentials. These states are virtually localized, so that their presence may make a QC less conductive.
\end{abstract}



\section{Introduction}

The one-electron properties of the Fibonacci lattice (FL) and the likes have been extensively investigated because they are considered to be one-dimensional (1D) models for quasicrystals (QCs); these lattices are quasiperiodic and called quasilattices (QLs). One of the remarkable features of these QLs is that all the one-electron eigenstates in them are neither extended nor localized in the standard meaning, but are of an intermediate kind. More precisely, these states exhibit multifractal properties and called {\it critical states}. The quantum dynamics of electrons associated with critical states is characterized by the anomalous quantum diffusion, which may be related to the unconventional transport properties of real QCs. \cite{BeMa93,Sire94,SchBaBell98,RoTrMa97,Mayou} However, it has been revealed recently that most of the 1D QLs whose one-electron properties have been investigated so far belong to a special class of QLs from the point of view of a general classification scheme of QLs, although the class is the simplest and, therefore, the most important. \cite{FuNi00,NiFu00} Moreover, it has also been reported that there exist new classes of QLs whose one-electron properties belong to a different universality class from that of the conventional class. \cite{FuNi00} The purpose of this paper is to investigate this subject in more detail.

Here, we shall summarize briefly the classification of 1D QLs. \cite{NiFu00} 1D QLs are in general generated by the cut-and-projection method \cite{DuKa85} from 2D periodic lattices, i.e., the mother lattices; each of the mother lattices has a hyperscaling symmetry \cite{Ni89} characterized by a quadratic irrational $\tau$ which assumes $(1 + \sqrt{5})/2$ (the golden ratio), $1 + \sqrt{2}$ (the silver ratio), etc. This class of QLs are naturally divided into distinct subclasses, each of which is specified by the relevant quadratic irrational. Different QLs in each subclass are specified by different windows which are used in the projection method. A QL with a generic window is ternary in the sense that there exist three types of lattice spacings, $s$, $m$, and $l$, where the three symbols refer to the words, ``short'', ``medium'', and ``long'', respectively. Each subclass is further divided into three finer subclasses, type I, type II, and type III, where the sizes of the windows for these subclasses are integral, rational, and irrational, respectively, with respect to the quadratic field $Q[\tau]$. \cite{NiFu00} Of these finer subclasses, the former two are of particular importance because they have inflation symmetries (self similarities). A generic type I QL for each $\tau$ is represented as a decoration of a binary self-similar QL; that is, all the type I QLs for each $\tau$ form a single mutual-local-derivability class (MLD class), \cite{Ba91} while the self-similar member can be chosen as the representative of the class. The FL and the likes which are binary are such representatives. On the other hand, the type II QLs, which are essentially ternary, are further classified into an infinite number MLD classes. \cite{NiFu00} Again, we can choose a self-similar QL as the representative of each MLD class, and other members are given as appropriate decorations of the representative.

The quadratic irrational characterizing a type I MLD class including the FL is the golden ratio, $\tau = (1 + \sqrt{5})/2$, while those for other type I MLD classes are other ``precious ratios (means)'', e.g., the silver ratio, $1 + \sqrt{2}$. Note that there exist an infinite series of type II MLD classes which are characterized by a single quadratic irrational.

Most aperiodic 1D structures generated by inflation rules (or, equivalently, substitution rules) are not QLs in our definition. The Thue-Morse sequence, the period doubling sequence, and the Rudin-Shapiro sequence are familiar examples. Although our definition of QLs is rather restrictive, the one-electron properties of type II QLs remained intact until a previous work by the present authors appeared. \cite{NiFu00} This work will be summarized shortly. It should be emphasized that type II QLs form an overwhelming majority in comparison with the conventional QLs, i. e., the type I QLs.

We can investigate the one-electron properties of a QL by the use of the Schr$\ddot{\rm o}$dinger equation in the tight-binding approximation:
\begin{eqnarray}
t_{j-1,j}\Psi_{j-1} + t_{j,j+1}\Psi_{j+1} = E\Psi_{j},
\label{schro.eq.}
\end{eqnarray}
with $t_{j,j+1}$ and $\Psi_j$ being the hopping integral and the probability amplitude, respectively, while $j$ runs over all the sites of the relevant QL. We have adopted the so-called bond model because we are interested in the properties which are independent of the microscopic details of the model. The geometry of a QL can be incorporated into Eq\verb/./(\ref{schro.eq.}) by assuming that $t_{j,j+1}$ takes one of the three values, $t_s$, $t_m$, and $t_l$, depending on the type of the relevant bond. The pioneering work of Kohmoto et al \cite{KKT83,KST87} showed that a similar model based on the FL exhibits quite unconventional one-electron properties; that is, 1) {\it the energy spectrum is purely singular-continuous or, equivalently, it has zero Lebesgue measure with no point component}, while 2) {\it all the eigenstates are neither extended nor localized but are critical}. Both the energy spectrum and the eigenfunctions exhibit multifractal properties. Note that critical states are distinctive from localized ones in that the eigenfunctions are not square-normalizable.

The integrated density of states (IDOS) per site, $H(E)$, is a fundamental quantity characterizing the one-electron properties of a macroscopically homogeneous 1D system. The scaling exponent $\alpha(E)$ is defined by the asymptotic behavior
\begin{eqnarray}
\Delta H &\equiv&
  H(E + \frac{1}{2}\epsilon) - H(E - \frac{1}{2}\epsilon)
 \sim \epsilon^{\alpha(E)},
\end{eqnarray}
with $E$ being the reference energy and $\epsilon$ a small parameter to be extrapolated to zero. \cite{KST87,HiKo92} It provides the local dimension of the energy spectrum at energy $E$ and satisfies the inequality $0 \le \alpha \le 1$. For the purely singular-continuous spectrum of the FL, the exponents $\alpha(E)$ distribute continuously within an interval, $\left[\alpha_{\rm min},\alpha_{\rm max}\right]$, with $0 < \alpha_{\rm min} < \alpha_{\rm max} < 1$. The distribution yields the $f(\alpha)$ spectrum of the energy spectrum. \cite{Ha86,Ko88} The three types of energy spectra in one dimension, namely, a point spectrum, an absolutely-continuous one, and a singular-continuous one, are associated with localized states, extended states, and critical states, respectively. The exponent $\alpha(E)$ gives a measure of the degree of localization of the eigenstate at energy level $E$; $\alpha(E)$ assumes $0$ if $E$ is on an isolated energy level, while it assumes $1$ if $E$ is inside an absolutely-continuous spectrum. Therefore, the spectral characterization by the scaling exponent $\alpha(E)$ is a basic problem in the studies of one-electron properties of QCs.

In order to analyze the asymptotic behavior of one-electron eigenstates in the FL, several renormalization-group (RG) methods have been developed, including the trace map, \cite{KKT83,KST87} the exact real-space RG, \cite{AsSt88} and the perturbational real-space RG methods. \cite{NiNo86} Although all these methods are independent in their formulations, they all rely on the inflation symmetry of the underlying structure. These methods have also been applied to general 1D structures with inflation symmetries. \cite{KoNo90,ChaKa89,BaLu94} One of the important consequences of these studies is that the one-electron properties of any decoration of the FL are characterized by the common kinds of fixed points to those for the FL. \cite{Ko92} In other words, all these QLs belong to a single universality class with respect to one-electron properties in the standard meaning of the RG theory. It follows then the eigenstates for all the QLs in a single type I MLD class exhibit power-law decay with fractional exponents as for the case of the FL. Most of the previous generalizations of the FL correspond to this universality class. \cite{Ho88,FuLiZhSr97}

Meanwhile, there has also been a large amount of work of mathematicians concerning the properties of the energy spectra of 1D systems with inflation symmetries. \cite{BoGh93} The analyses have extended over a wide range of 1D structures surpassing the list of type I QLs. \cite{foot} In almost all cases the energy spectra have turned out to be rigorously singular-continuous, confirming that the one-electron eigenstates in these self-similar structures are critical. Note, however, these results do not specify the detailed properties of the critical states, so that there still remains a possibility of different universality classes connected with some self-similar structures from that of the type I QLs.

The present authors have recently reported some preliminary results on the one-electron properties of several type II QLs. Indeed, their numerical results on the $f(\alpha)$ spectra suggest strongly that all the eigenstates are critical. But they have also found that there exists a certain unconventional type of critical states called {\it marginal critical states}. With the perturbational real-space RG method, they have proved that the asymptotic behavior of these new states is described by a stretched exponential decay, where the corresponding scaling exponent $\alpha$ vanishes. The decay is much stronger than power-law decay for conventional critical states, and marginal critical states can be thought of as special critical states which are almost localized. This result is closely connected with the self-similarity of the type II QLs, and it proves that type II QLs belong to different universality classes from that of the type I QLs. We present in this paper a full account of the presence of marginal critical states in type II QLs based on the exact real-space RG analysis.

This paper is organized as follows. In Sec\verb/./\ref{sec:structure}, we present several examples of type II QLs which are discovered by the authors. \cite{NiFu00} Then numerical results for the properties of the energy spectrum of a type II QL are presented in Sec\verb/./\ref{sec:numerical}, and the asymptotic behavior of the one-electron eigenstates are analyzed in Sec\verb/./\ref{sec:scaling} by the use of the exact real-space RG scheme. Sec\verb/./\ref{sec:discussions} is devoted to discussions about related subjects and to the conclusion of the paper.

\section{Type II QLs in one dimension}
\label{sec:structure}

We list in Table \ref{table1} representatives of six MLD classes of 1D QLs: these MLD classes, (A) to (F), have been investigated in our recent work. \cite{NiFu00} The inflation rule for each of the six representative QLs is listed in the last column. Of the six representatives, only the one given in the first row is of type I. Note that the first three QLs are obtained from a common mother lattice but with different windows. Similarly, the mother lattice used to obtain QL(E) is the same as that for the FL, but the window is different. We have used the symbol ``QL(E)'' to denote the QL listed in the fifth row of Table \ref{table1}. Remember that the spacing $s$ is isolated in all the QLs in Table \ref{table1} but QL(D).

In order to illustrate some of the features of these QLs, we shall investigate in more detail QL(B). The three spacings of the QL are given by $s = 1$, $m = \sqrt{2}$, and $l = 1 + \sqrt{2}$ with an appropriate unit of length. The inflation rule for this QL is characterized by the inflation matrix
\begin{equation}
{\bf M} = \left( \begin{array}{ccc}
1 & 1 & 1\\
1 & 0 & 0\\
0 & 1 & 2\\
\end{array} \right), \label{eqn:mld19b}
\end{equation}
whose Frobenius eigenvalue is equal to $\sigma=1+\sqrt{2} \; (=\tau)$. The right Frobenius eigenvector is proportional to the row vector $(s\verb/,/\,m\verb/,/\,l)$, while the three components of the left Frobenius eigenvector, $^t(\tau^{-1}\verb/,/\tau^{-2}\verb/,/\tau^{-1})$, are proportional to the frequencies of the three spacings in the QL, where the left superscript ``$t$'' stands for the transpose operation. It happens that two spacings $s$ and $l$ have an equal frequency; this is the case for QL(C), QL(E), and QL(F) as well but not for QL(D). It follows that the average lattice spacing of QL(B) is given by $\bar{a}=4\tau^{-1}$.

If the inflation operation for QL(B) is repeated $n$ times, we obtain the spacings of the $n$-th generation, $s^{(n)}$, $m^{(n)}$, and $l^{(n)}$. For instance, the first several generations of $s^{(n)}$ are: $s^{(0)}=s$, $s^{(1)}=ms$, $s^{(2)}=lsms$, $s^{(3)}=llsmslsms$, $\cdots$. The numbers of the original three types of spacings, $s$, $m$, and $l$, included in those of the $n$-th generation are given by the elements of the matrix ${\bf M}^n$. Using this result, we can show that the total number of spacings included in $s^{(n)}$ are given by $L_n=(K_{n+1}+1)/2$, where $K_n$ is determined recursively by the equation $K_{n+1}=2 K_n + K_{n-1}$ with the initial conditions $K_0=K_1=1$. Specifically, $L_n$ yields a series of numbers: 1, 2, 4, 9, 21, 50, 120, 289, 697, 1682, 4060, $\cdots$. It is evident that $L_{n+1}/L_{n}$ and $K_{n+1}/K_{n}$ tend both to $\sigma=\tau=1+\sqrt{2}$ as $n$ goes to infinity. $s^{(n)}$, $m^{(n)}$, or $l^{(n)}$ provides us with the $n$-th periodic approximants of QL(B). The length of $s^{(n)}$ is asymptotically given by $L_n \bar{a}$, so that we need not distinguish it from $L_n$ if the unit cell is so chosen that $\bar{a}=1$.

\section{Numerical results}\label{sec:numerical}

\begin{figure}
\begin{center}\epsfile{file=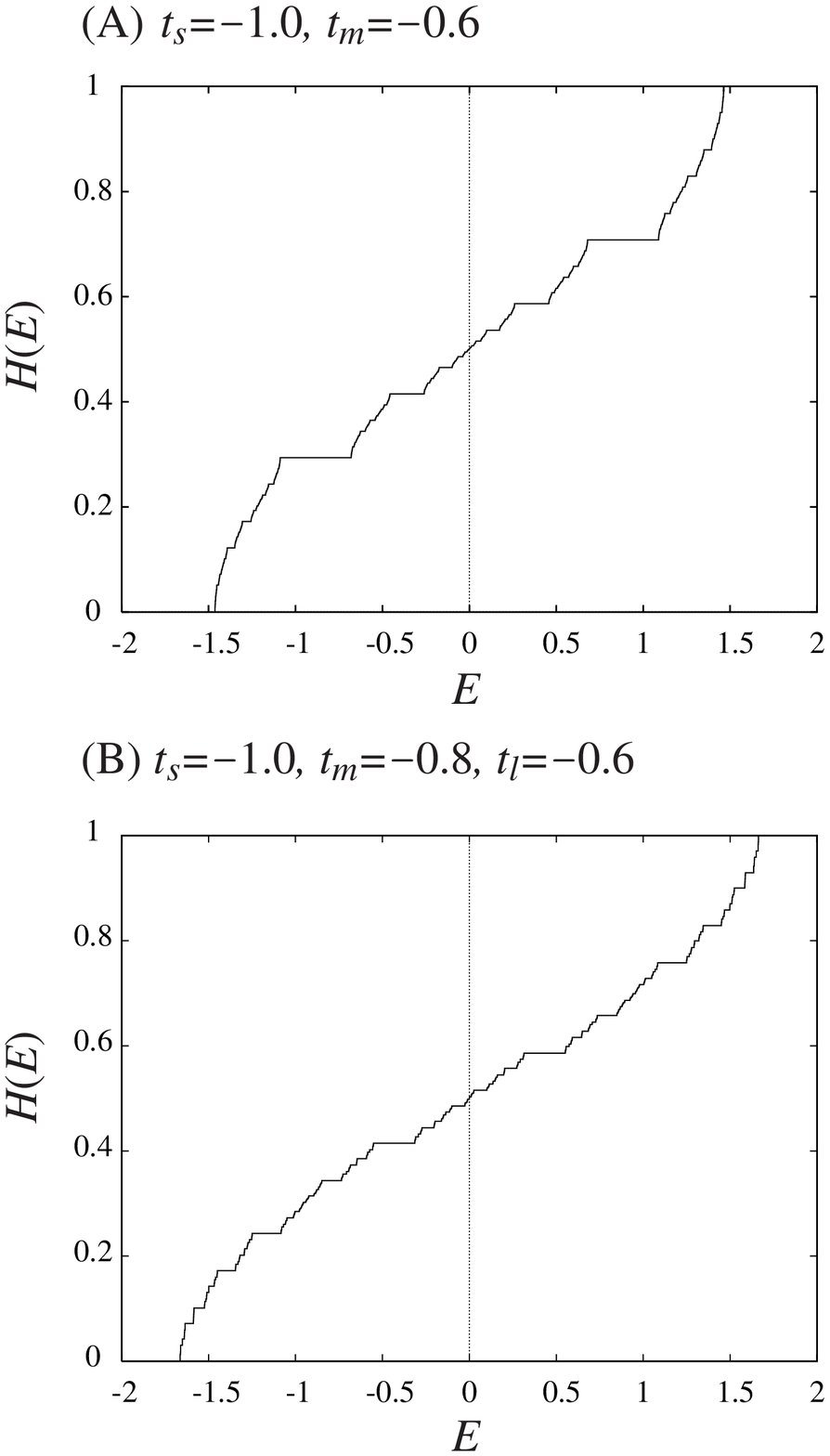,height=17cm}\end{center}
\caption{The IDOSs of QL(A) (top) and QL(B) (bottom). The values of the hopping integrals taken are indicated in the figures.}
\label{IDOS}
\end{figure}

We investigate the energy spectrum of QL(B). Since there is no analytical method for exact evaluation of the energy spectrum, we resort to numerical calculations for periodic approximants of the QL. The IDOS obtained numerically is shown in Fig\verb/./\ref{IDOS}. Also the IDOS of QL(A) is shown for comparison in the same figure. Both the IDOSs exhibit devil's staircases because the spectra are purely singular-continuous.

A singular-continuous energy spectrum includes an infinite number of gaps with varying sizes. We shall call an energy level an edge level if it is located on an end of a gap. Energy levels accumulate at an edge level from one side only, so that the bottom level and the top one are regarded to be edge levels as well. An eigenstate is called an edge state if its energy coincides an edge level. Edge states will be found to be key states to understand the one-electron properties of type II QLs.

\begin{figure}
\begin{center}\epsfile{file=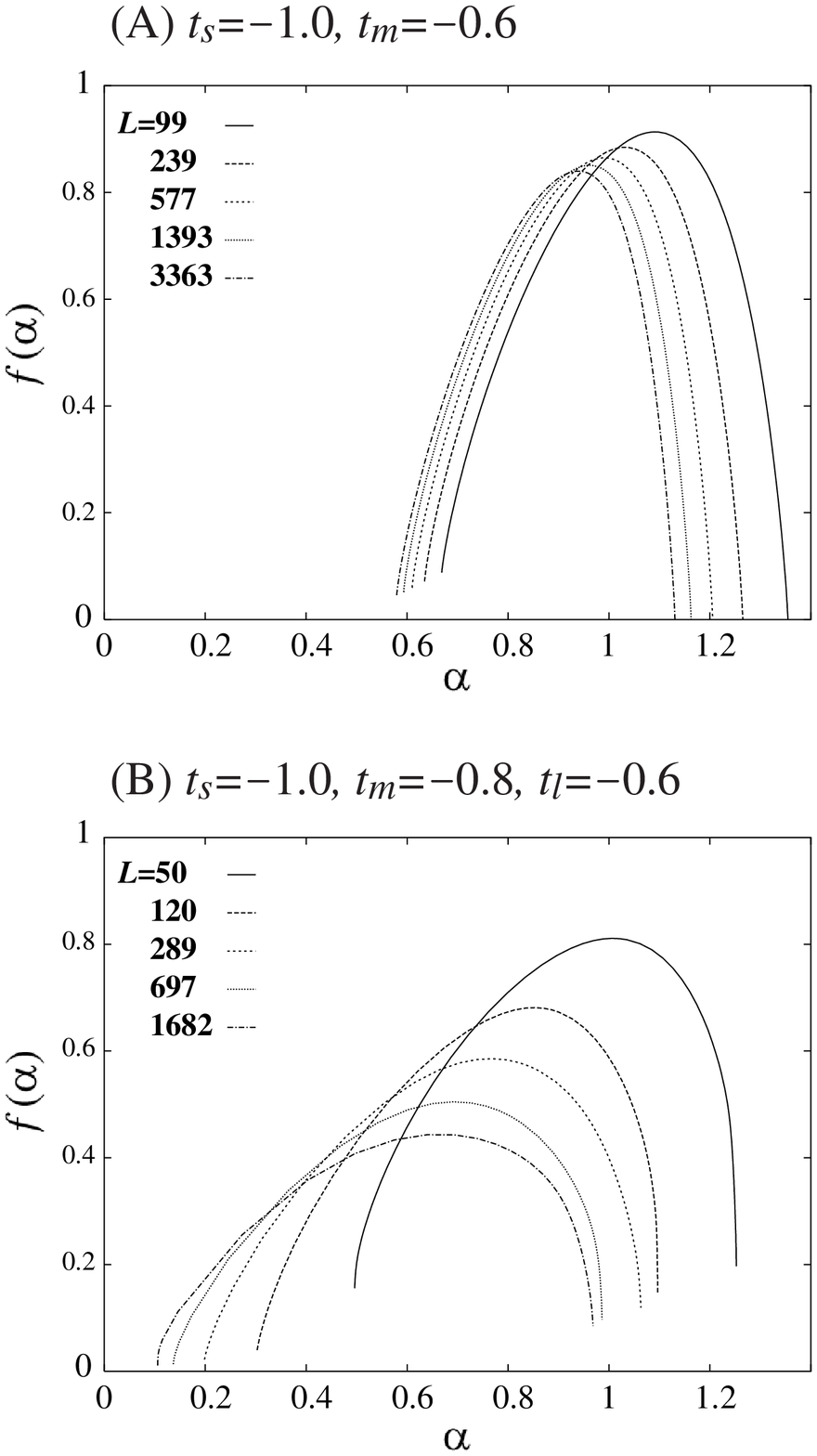,height=17cm}\end{center}
\caption{The $f(\alpha)$ spectra of the energy spectra for periodic approximants of QL(A) (top) and QL(B) (bottom). The hopping integrals as well as the sizes $L$ of the periodic approximants are given in the figures, where we mean by a size of a periodic approximant the number of sites in its unit cell. $f(\alpha)$ spectra in the thermodynamic limit are to be obtained from these data with the finite-size scaling analysis but we do not show them because the results do not have sufficient accuracy.}
\label{Falpha}
\end{figure}

The multifractal analysis of an energy spectrum can be performed with the formalism of Kohmoto. \cite{Ko88} The $f(\alpha)$ spectra calculated for approximants of several generations for the two kinds of QLs, QL(A) and QL(B), are shown in Fig\verb/./\ref{Falpha}. The left end of the support of the $f(\alpha)$ spectrum for QL(B) seems to tend, in the thermodynamic limit, to the lower end of the $\alpha$-axis, implying that some of the eigenstates in this QL tend to be localized. This is surprising because the distribution of the hopping integrals is narrower in QL(B) than in QL(A).

\begin{figure}
\begin{center}\epsfile{file=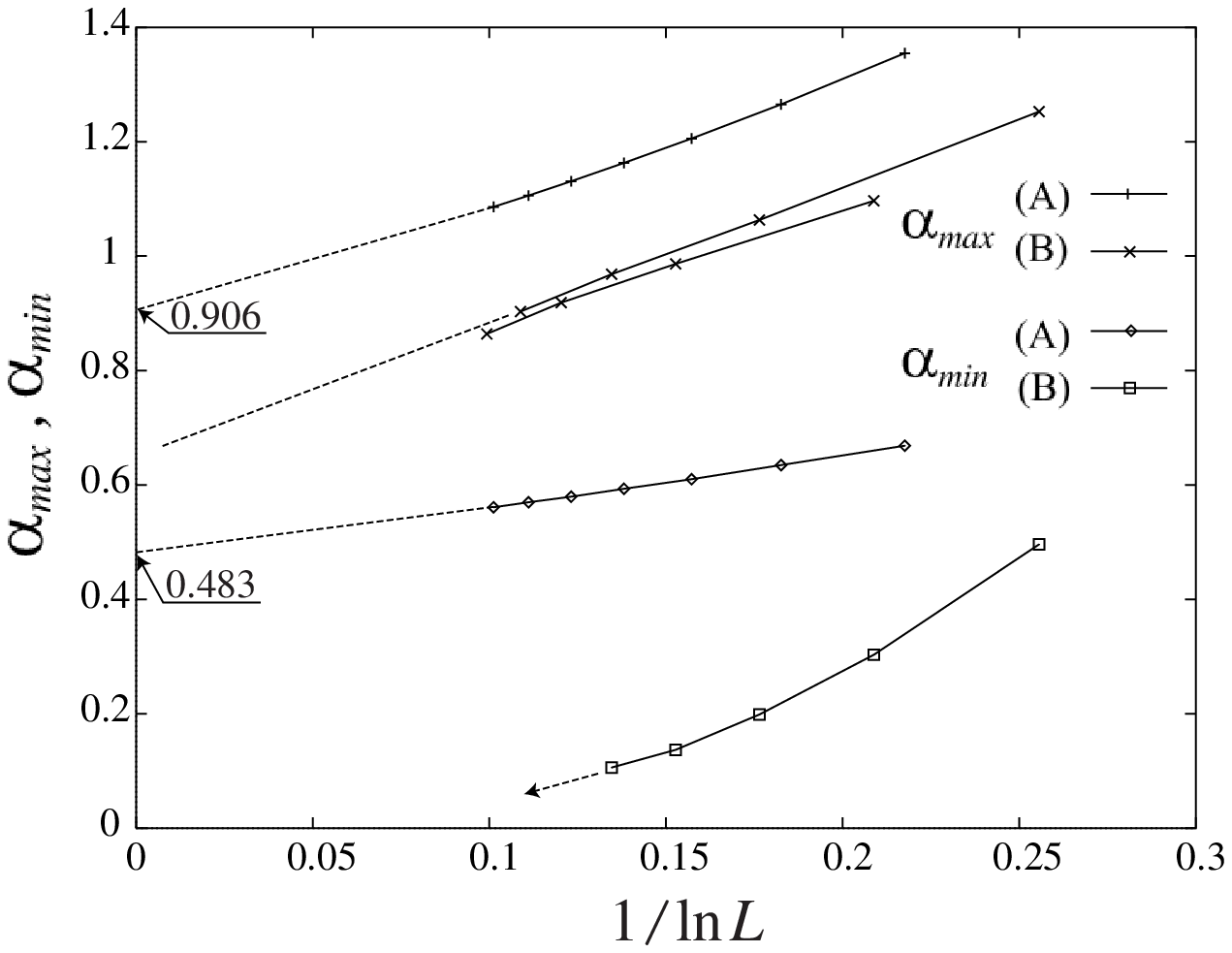,height=8cm}\end{center}
\caption{Finite-size scaling plots of $\alpha_{min}$ and $\alpha_{max}$ for QL(A) and QL(B) with the same values of hopping integrals as in Fig. \hspace{-3pt}2. The abscissa is $1/\ln{L}$, where $L$ is the number of sites in the unit cell.}\label{aminamax}
\end{figure}

We show in Fig\verb/./\ref{aminamax} finite-size scaling plots of $\alpha_{min}$ and $\alpha_{max}$. If a quantity obeys power-law scaling, the relevant plot exhibits linear behavior. We can obtain the thermodynamic limits of these exponents from the plots when they obey power-law scaling. Our estimates are: $\alpha_{min}=0.483$ and $\alpha_{max}=0.906$ for QL(A). For QL(B), $\alpha_{min}$ does not obey power-law scaling and seems to vanish in the thermodynamic limit, while $\alpha_{max}$ exhibits two-cycle and power-law scaling seems to be preserved during first several steps, where $\alpha_{max}$ in the thermodynamic limit is estimated to be $0.6\sim 0.7$ with only one digit of accuracy.

Similar numerical analyses for other type II QLs in Table \ref{table1} strongly indicate that the vanishing of $\alpha_{min}$ is a universal feature of type II QLs, being intrinsically different from the case of type I QLs. It is also observed numerically that the energy spectrum of a type II QL exhibits non-power-law scaling at an edge level, suggesting that the local scaling exponent $\alpha$ for the edge level vanishes. In the next section, the scaling exponent $\alpha$ for an edge level of a type II QL is shown to be exactly zero, providing us with a direct proof that $\alpha_{min}=0$.

\section{A real-space RG analysis}
\label{sec:scaling}

The authors have investigated in a previous paper \cite{FuNi00} (to be referred to as FN) the edge states of a type II QL by means of a real-space RG method. \cite{NiNo86,BaLu94} This method is, however, based on the perturbation theory, and applies only to the strong modulation regime, where $|t_s|\gg |t_m|\gg |t_l|$. In this paper we shall employ an exact real-space RG method which has been previously applied to type I QLs. \cite{AsSt88,ChaKa89} This method is based on the Dyson equation for the one-electron Green function in real space. Since the method does not rely on any approximation, it is applicable to more general situations, including the weak modulation regime.

\begin{figure}
\begin{center}\epsfile{file=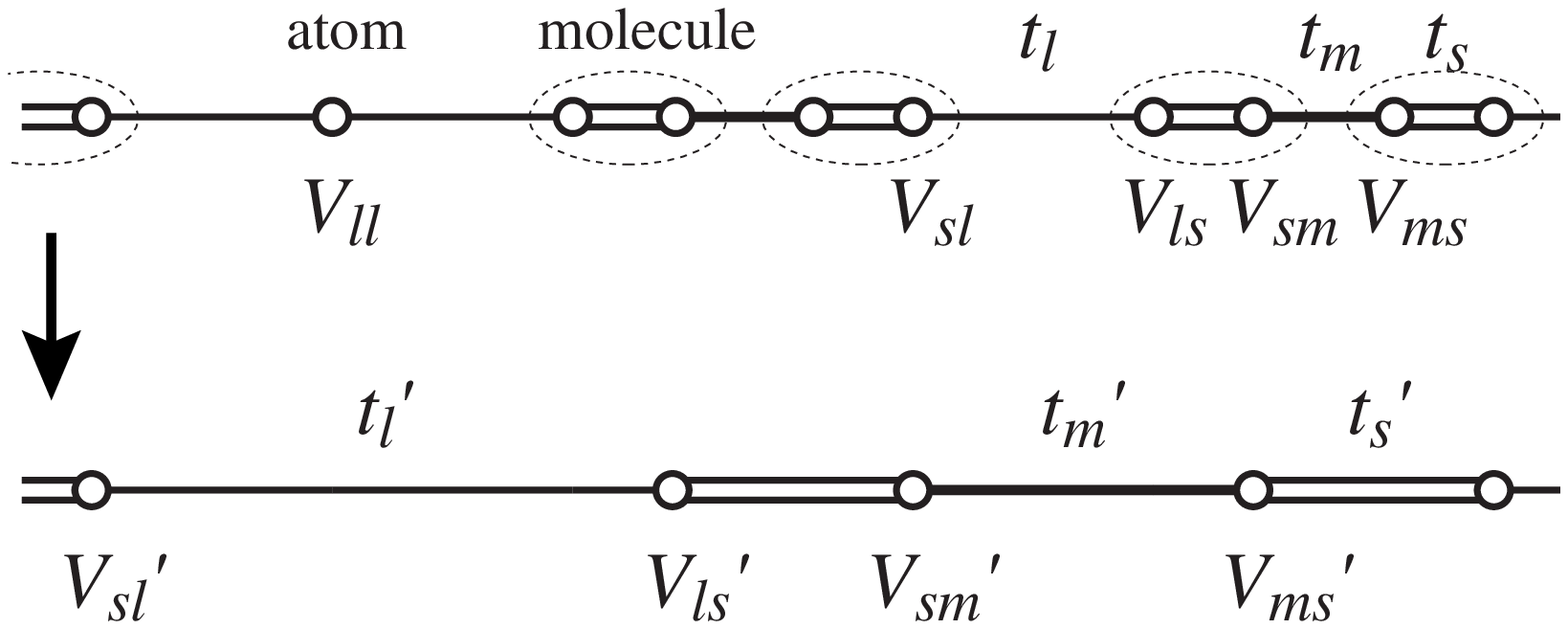,height=4.3cm}\end{center}
\caption{The decimation procedure appropriate for the bottom level of QL(B).}\label{fig:coarse}
\end{figure}

We shall illustrate the theory with its application to QL(B). The decimation appropriate for the bottom energy level of this QL is shown in Fig\verb/./\ref{fig:coarse}, where all the sites except those on the right side of the type $s$ spacings are decimated; the decimation conforms to the inflation rule of the QL. This decimation is applicable to any edge level as well. The relevant one-electron Hamiltonian is renormalized at each step of the decimation procedure, and, at the $n$-th step, there appear three types of effective hopping integrals, $t_{s}^{(n)}$, $t_{m}^{(n)}$, and $t_{l}^{(n)}$, and five types of effective site potentials, $V_{sm}^{(n)}$, $V_{ms}^{(n)}$, $V_{ll}^{(n)}$, $V_{sl}^{(n)}$, and $V_{ls}^{(n)}$, which are associated with the five types of local environments of the QL. \cite{NiFu00} The initial values of these eight-parameters are given by the bare parameters, $t_x^{(0)} = t_x$ and $V_{xy}^{(0)}\equiv 0$ ($x,y = s,m,l$). The two sets of effective parameters corresponding to two successive steps of the decimation are related to each other by
\begin{eqnarray}
&&t'_{s} = \frac{t_{m}t_{s}}{E-V_{ms}},\;
  t'_{m} = \frac{t_{l}t_{s}}{E-V_{ls}},\;
  t'_{l} = \frac{t_{l}t'_{m}}{E-X},\nonumber\\
&&V'_{sm} = V_{sl}+\frac{t_{s}^2}{E-V_{ms}}+\frac{t_{l}^2}{E-V_{ls}},\nonumber\\
&&V'_{ms} = V_{sm}+\frac{t_{s}^2}{E-V_{ls}}+\frac{t_{m}^2}{E-V_{ms}},\nonumber\\
&&V'_{ll} = V_{sl}+\frac{t_{s}^2}{E-V_{ls}}+\frac{t'^2_{m}}{E-X}+\frac{t_{l}^2}{E-X},\label{eq:rg}\\
&&V'_{sl} = V_{sl}+\frac{t_{s}^2}{E-V_{ms}}+\frac{t_{l}^2}{E-X},\nonumber\\
&&V'_{ls} = V_{sm}+\frac{t_{s}^2}{E-V_{ls}}+\frac{t'^2_{m}}{E-X}+\frac{t_{m}^2}{E-V_{ms}}\nonumber
\end{eqnarray}
with $X = V_{ll}+{t_{l}^2}/{(E-V_{ls})}$. Note that the primed parameter, $t'_{m}$, participates in the right hand sides of some of the equations. This 8D map is transformed with the seven dimensionless parameters,
\begin{eqnarray}
&&a = \frac{E-V_{sm}}{t_s},\;
  b = \frac{E-V_{ms}}{t_s},\;
  c = \frac{E-V_{ll}}{t_s},\;
  d = \frac{E-V_{sl}}{t_s},\nonumber\\
&&e = \frac{E-V_{ls}}{t_s},\;
  f = \frac{t_m}{t_s},\;
  g = \frac{t_l}{t_m},\label{eq:sevenpara}
\end{eqnarray}
into a 7D map:
\begin{eqnarray}
&& a'  =   -\frac{e - b d e + b f^2 g^2}{e f},\nonumber\\
&& b'  =   \frac{-b + a b e - e f^2}{e f},\nonumber\\
&& c'  =   \frac{b (c - c d e + d f^2 g^2 + e f^2 g^2)}{f (- c e + f^2 g^2)},\nonumber\\
&& d'  =   \frac{c e - b c d e - f^2 g^2 + b d f^2 g^2 + b e f^2 g^2}{f (- c e + f^2 g^2)},\label{eq:7dmap}\\
&& e'  =   \frac{b c - a b c e + c e f^2 + a b f^2 g^2 - f^4 g^2}{f (- ce + f^2 g^2)},\nonumber\\
&& f'  =   \frac{b g}{e},\nonumber\\
&& g'  =   -\frac{e f g}{- c e + f^2 g^2}.\nonumber
\end{eqnarray}
A point of the relevant 7D space is represented by the 7D vector: ${\bf a} = (a, b, c, d, e, f, g)$.

\begin{figure}
\begin{center}\epsfile{file=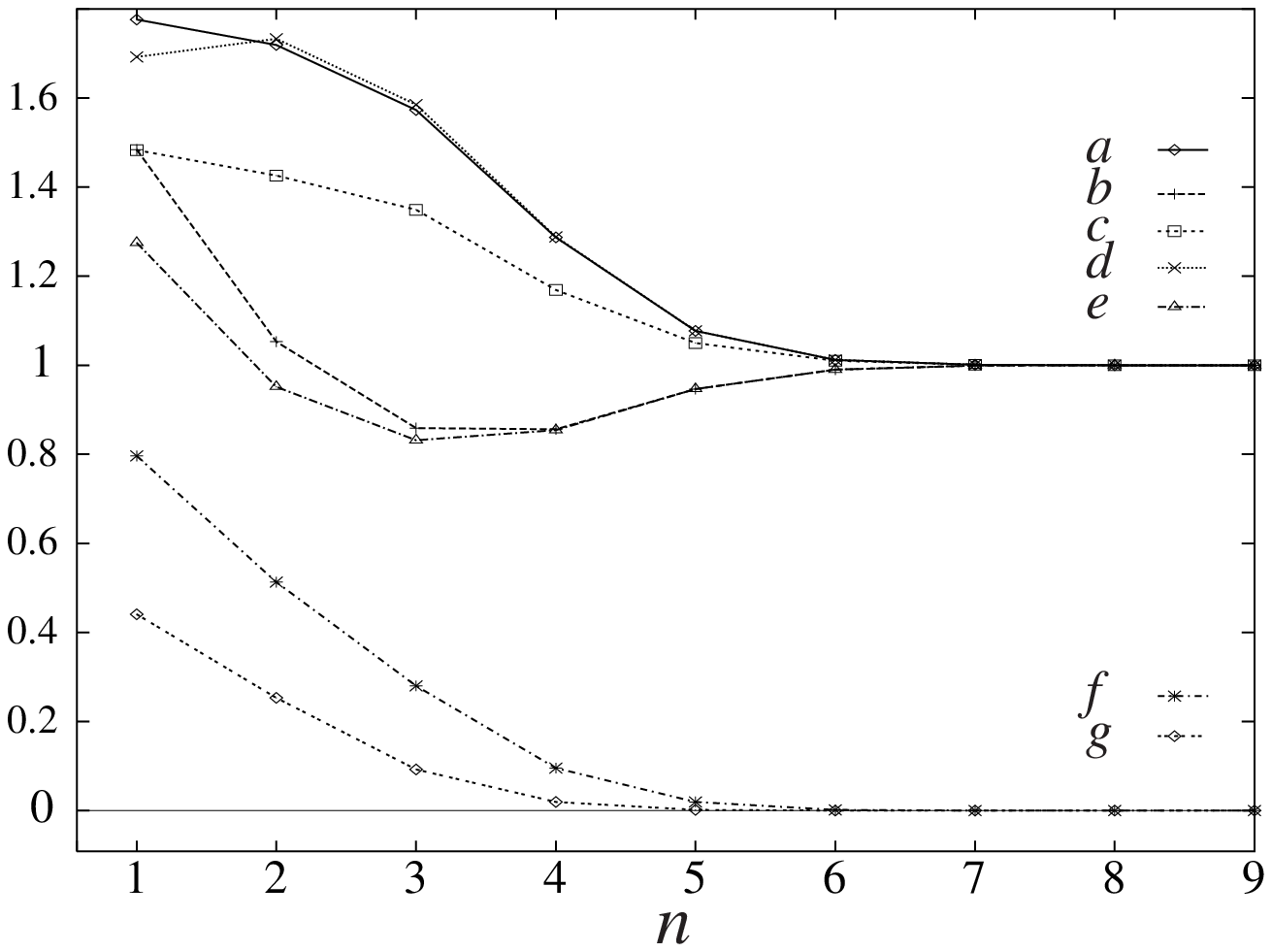,height=7.8cm}\end{center}
\caption{The flow of the parameters $a, b, c, d, e, f, g$ is plotted against $n$, the generation number, for $t_s = -1.0, t_m = -0.8, t_l = -0.6$.}
\label{fig:Ai}
\end{figure}

The 7D map yields a sensible result only when it has a fixed point. This is the case when $E$ is tuned to an edge level. Since the edge levels cannot be estimated analytically, we have to obtain them numerically. We take the bottom level as the representative of edge levels. Its value can be evaluated accurately enough on the basis of a finite-size scaling analysis. The RG flow calculated numerically for the bottom level is shown in Fig\verb/./\ref{fig:Ai} up to the $9$-th step. We find that the flow converges to ${\bf a}_{I} = (1,1,1,1,1,0,0)$, which is a fixed point of the 7D map. We can conclude from this result and Eq\verb/./(\ref{eq:sevenpara}) that five quantities of the type $E-V_{xy}$ ($x,y = s,m,l$) are all equal to $t_s$ at the fixed point, while $t_m$ and $t_l$ can be ignored in comparison with $t_s$. Therefore, the quasiperiodic chains of atoms are separated eventually into isolated ``atoms'' and diatomic ``molecules'' which are combined through the transfer integral $t_s$. That is, the fixed point belongs to the limit of the strong modulation regime. This means that RG flow brings the system into the strong modulation regime even if the system is, at the beginning, in a weak modulation regime. It follows that the perturbational RG theory is applicable in a vicinity of the fixed point, and main conclusions in FN are justified for the weak modulation regime as well. Incidentally, we may conclude from Eqs.(\ref{eq:rg}) and (\ref{eq:7dmap}) that $a\simeq d$ and $b\simeq e$ in a close vicinity of the fixed point ${\bf a}_{I}$ (see Fig\verb/./\ref{fig:Ai}), since all the denominators in Eq.(\ref{eq:rg}) are of the same order as $t_s$ and both $t_m$ and $t_l$ are negligible in comparison with $t_s$.

\begin{figure}
\begin{center}\epsfile{file=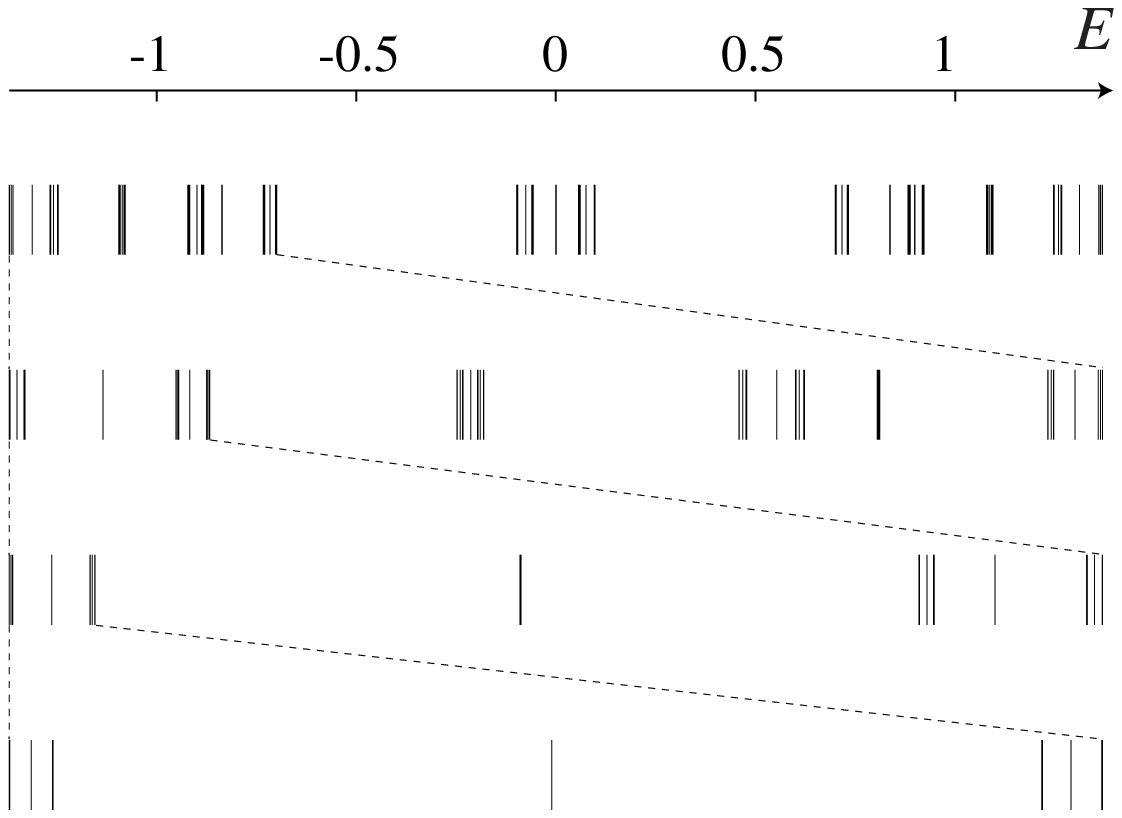,height=7.5cm}\end{center}
\caption{The symmetric triplet structure of the energy spectrum of QL(B), where the hopping integrals are taken to be $t_s=-1.0$, $t_m=-0.5$, and $t_l=-0.3$. When the leftmost subband is expanded repeatedly, the triplet structure shrinks, since the ``atoms'' and ``molecules'' are isolated in the relevant limit.}
\label{fig:spec}
\end{figure}

A decimation procedure in the real space RG theory is nothing but a coarse-graining in the real space. The same procedure effects as a magnification in the energy space. It follows that the energy spectrum has a symmetric triplet structure in a vicinity of each edge level; the triplet is composed of three bands. The central band of the triplet is derived from isolated ``atoms'', while the two satellite bands from isolated ``molecules'', which have bonding and anti-bonding states. This has been numerically confirmed as shown in Fig\verb/./\ref{fig:spec} (see also Fig\verb/./2 in FN). The relative weights of the three subbands are determined by the frequencies of the atoms and the molecules appearing in the system to be $\tau^{-1}:\tau^{-2}:\tau^{-1}$ in the order of them in the spectrum, where $\tau = 1+\sqrt{2}$.

Let us consider an asymptotic flow in the vicinity of the fixed point of the present RG. In a neighborhood of the fixed point of the 8D map, we can approximately replace by $t_{s}$ the denominators of the three equations in the first row of Eq\verb/./(\ref{eq:rg}), so that we obtain
\begin{equation}
  t'_{s} = \frac{t_{m}t_{s}}{t_{s}},\;
  t'_{m} = \frac{t_{l}t_{s}}{t_{s}},\;
  t'_{l} = \frac{t_{l}^2t_{s}}{t_{s}^2}.
  \label{eq:sbrg}\\
\end{equation}
That is, the relevant 3D subspace of the 8D space is closed approximately in that neighborhood. Remember that the structures of the numerators in the right hand sides of the three equations above are parallel to those of the inflation rule of QL(B) (cf. Table \ref{table1}). The 3D map above is reduced to the 2D map:
\begin{eqnarray}
&& \left\{ \begin{array}{ll}
f' = & g,\\
g' = & fg,
\end{array}\right.\label{eq:2Dmap}
\end{eqnarray}
whose standard linearization in a neighborhood of the fixed point $(0,0)$ yields
\begin{eqnarray}
&& \left(\begin{array}{c}
f' \\ g'\end{array}\right)
 = \left(\begin{array}{c}
g  \\ 0 \end{array}\right)
 =  \left(\begin{array}{cc}
0 & 1\\ 0 & 0\end{array}\right)
\left(\begin{array}{c}
f \\ g\end{array}\right).\label{eq:linear1}
\end{eqnarray}
The $2\times 2$ matrix at the right hand side is nilpotent because its square vanishes, and it cannot be diagonalized. This gives rise to a non-scaling behavior of the map at the fixed point. If new variables are introduced by $\hat{f} = \ln{f}$ and $\hat{g} = \ln{g}$, the non-linear map (\ref{eq:2Dmap}) can be changed into the linear map:
\begin{eqnarray}
&& \left( \begin{array}{c}\hat{f}' \\ \hat{g}' \\
          \end{array}\right) = {\bf M}
   \left( \begin{array}{c}\hat{f} \\ \hat{g}\\
          \end{array}\right), \quad {\bf M} =
   \left( \begin{array}{cc}0 & 1\\ 1 & 1\\
          \end{array}\right).\label{eq:linear2}
\end{eqnarray}
The matrix ${\bf M}$ can now be diagonalized, and the $n$-th member, $(\hat{f}_n, \;\hat{g}_n)$, of the orbit generated by the map has the asymptotic expression: $\hat{f}_n \approx \hat{g}_{n-1}\propto \rho^{n}$, where $\rho = (1+\sqrt{5})/2$ is the leading eigenvalue of ${\bf M}$; it happens that $\rho$ coincides the golden ratio. Finally, we obtain the asymptotic expressions for $f$ and $g$:
\begin{equation}
f_{n}\; = g_{n-1}\;
\propto \exp{[-c\rho^{n}]},
\label{asymptotic}\end{equation}
with $c$ being a positive constant, which depends on the system parameters through the initial condition of the original 8D map.

\begin{figure}
\begin{center}\epsfile{file=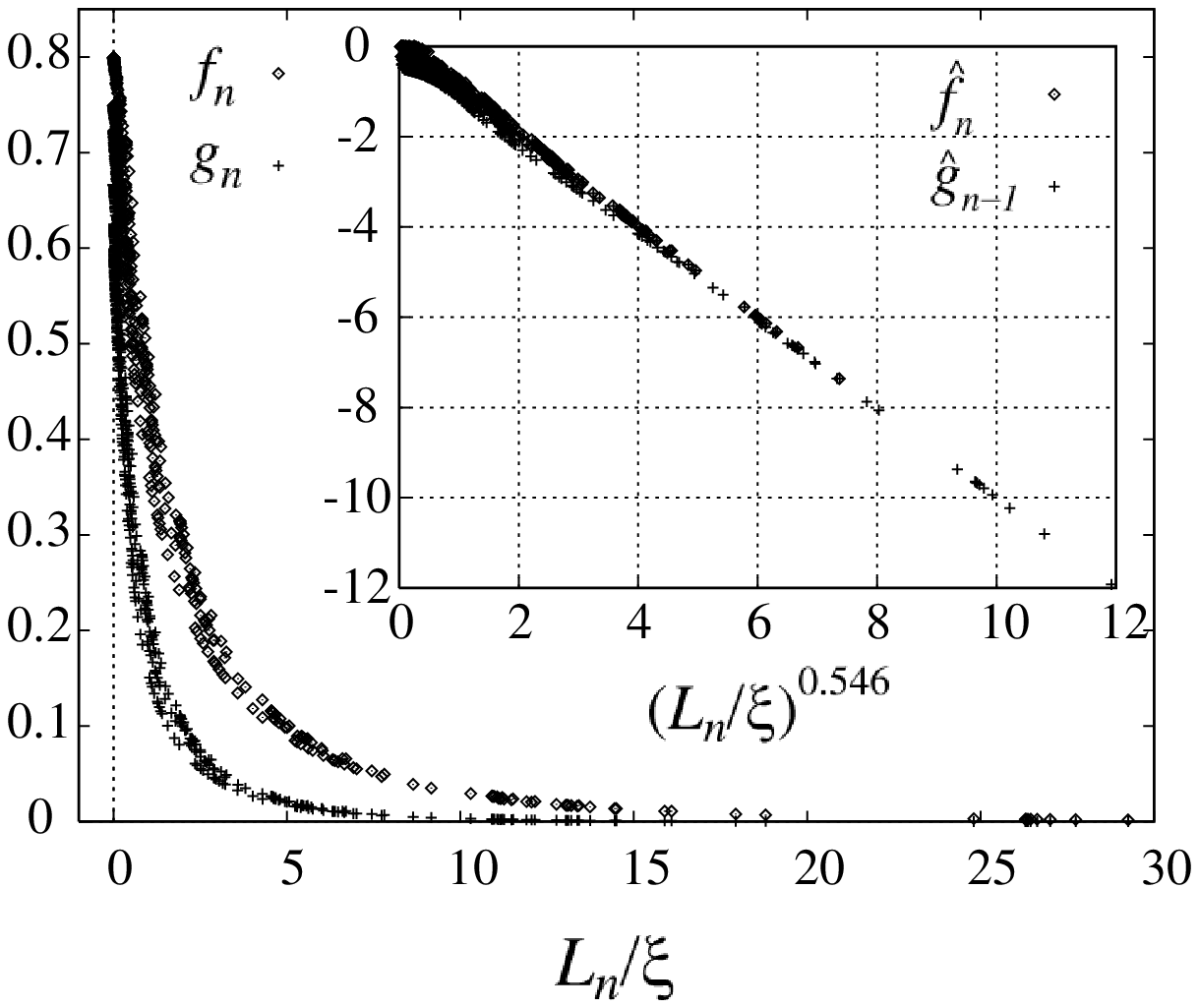,height=8cm}\end{center}
\caption{Numerical data of $f_n$ and $g_n$ ($t_s=-1.0$, $t_m=-0.9\sim -0.998$, $t_l=-0.8\sim -0.996$, $n=1\sim 7$) are plotted against $L_n/\xi$ with $\xi$ being evaluated through finite-size scaling analyses of $f_n$ for fixed system parameters. In the inset, the corresponding data of $\hat{f}_n$ and $\hat{g}_{n-1}$ are plotted against $(L_n/\xi)^{\nu}$ with $\nu=0.546$, confirming that $\hat{f}_{n}=\hat{g}_{n-1}\approx -(L_n/\xi)^\nu$.}\label{fig:scaleD}
\end{figure}

The ratio of the self similarity of the QL under consideration is equal to $\sigma = 1+\sqrt{2}$, and the length scale $L=L_n$ characterizing the $n$-th step of the renormalization transformation is proportional to $\sigma^{n}$. Therefore, the asymptotic expression, Eq\verb/./(\ref{asymptotic}), is rewritten as a stretched exponential:
\begin{equation}
f_{n} \propto  \exp{[-(L/\xi)^\nu]},
\end{equation}
where $\xi$ is a characteristic length depending on the system parameters and the exponent $\nu$ of the stretched exponential is given by
\begin{equation}
\nu=\ln{\rho}/\ln{\sigma}.
\end{equation}
To be more specific, we obtain $\nu=\ln{\left[\frac{1}{2}(1+\sqrt{5})\right]}/\ln{\left[1+\sqrt{2}\right]}\approx 0.5460$ for QL(B). The prediction that the asymptotic flow of the RG is characterized by a stretched exponential is confirmed numerically as shown in Fig\verb/./\ref{fig:scaleD}.

If the linearized form of Eq\verb/./(\ref{eq:2Dmap}) had a diagonalizable transformation matrix, the asymptotic flow would obey power-law, $L^{-1/\alpha}$, with $\alpha$ being the scaling exponent for the relevant energy level $E$; this is the case for the fixed point associated with a conventional critical state appearing in the case of a type I QL. On the other hand, a disordered system in an Anderson localization regime will exhibits an exponential law like $\exp{(-L/\xi)}$ with $\xi$ being the localization length of an exponentially localized eigenfunction. The one-electron eigenstates characterized by stretched exponentials are called in FN {\it marginal critical states}.

\section{Discussions and the conclusion}\label{sec:discussions}

In the preceding section, our analysis has been focused on QL(B) but a similar analysis is possible for each of the other four type II QLs in Table \ref{table1} as well. Again, the inflation rule provides the proper decimation procedure for the edge states, and the asymptotic behavior of the flow is governed by a map reduced into a similar 3D subspace of the 8D parameter space. The parallelism between the 3D map and the inflation rule of the relevant QL holds as well. The relevant 2D maps and some other quantities of all the type II QLs in Table \ref{table1} are listed in Table \ref{tab:dynmap}. The exponents $\nu$ of the stretched exponentials are not universal among different type II QLs. The value of $\nu$ for QL(D) happens to be the same as that for QL(E) and so does that of the scaling parameter $\sigma$, but this does not mean that the two QLs belong to a common universality class with respect to the one-electron properties because the 2D maps are distinct.

It is known that different QLs derived from a common mother lattice but with different windows have a common Fourier module which characterizes the positions of the peaks of the Dirac measures. It is surprising, therefore, that QL(A), QL(B), and QL(C), having a common mother lattice, belong to different universality classes with respect to the one-electron properties. Remind in this respect that the mother lattice is  common between the FL and QL(E) as well.

\begin{figure}
\begin{center}\epsfile{file=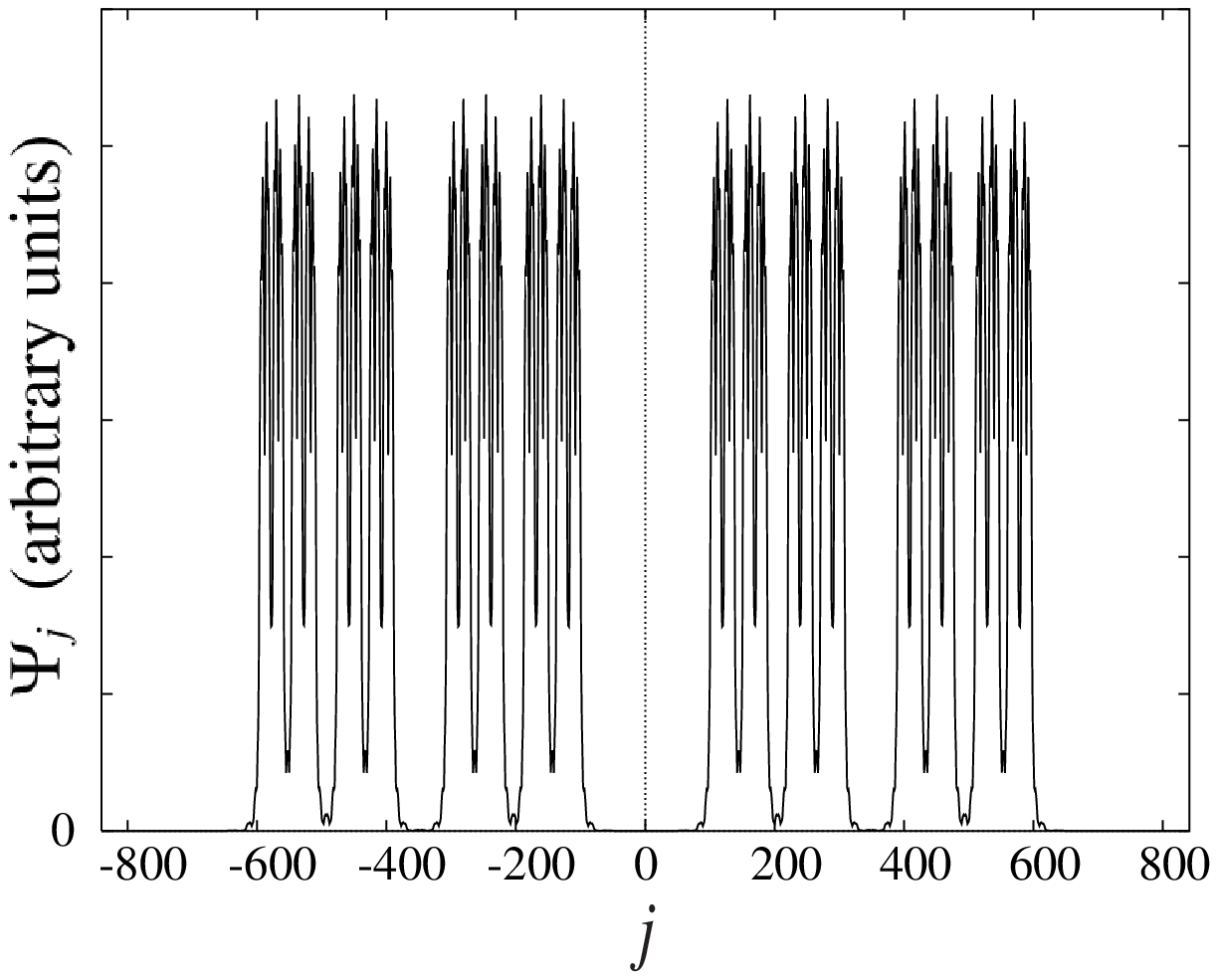,height=8cm}\end{center}
\caption{The eigenfunction of the bottom level of QL(B) in an approximant of $L=1682$ with $t_s=-1.0$, $t_m=-0.8$, and $t_l=-0.6$.}
\label{fig:wf}
\end{figure}

The presence of a non-trivial length scale $\xi$ makes a marginal critical state lack multifractality. This can be observed in the eigenfunction of the bottom state shown in Fig. \ref{fig:wf}. The whole eigenfunction can be divided into two identical parts, which can be divided further into two almost identical parts and so on. This hierarchical structure is connected with the structure of the RG near the relevant fixed point; a ``molecule'' formed by two ``atoms'' of each generation turns out to be an ``atom'' of the next generation. The weakening of the interatomic coupling with the generation as a function of the relevant scale in real space is given by a stretched exponential, so that the eigenfunction lacks self similarity. The coupling virtually vanishes at the generations where the length scale $L_n$ satisfies the inequality $L_n \gg \xi$, so that $\xi$ gives a measure of the scale of ``molecules'' which carry virtually localized wave functions. The estimated value of $\xi$ for the bottom state shown in Fig\verb/./\ref{fig:wf} is $\xi\sim 3$, while the size $L$ of an almost localized molecule in the figure is one for which $(L/\xi)^{\nu}$ exceeds the order of ten. The presence of marginal critical states makes the left edge, $\alpha_{min}$, of the support of the $f(\alpha)$ spectrum extend as far as the lower limit $\alpha = 0$. Since the whole distribution of $\alpha$ is shifted significantly downwards in this case, the one-electron eigenstates in a type II QL have strong tendency of localization. This may make a QC less conductive if its structure is described by a type II QL.

We shall investigate the relation between type II QLs and the so called {\it circle sequences} (CSs). \cite{AuGoLu88} We begin with introducing CSs. A CS is an infinite string of two digits, $0$ and $1$. It is specified by two fractional parameters, $\Delta, \zeta \in (0,1)$, with $\zeta$ being an irrational, and its $j$-th digit with $j \in {\bf Z}$ is given by $[j\zeta]-[j\zeta-\Delta]$ with $[\cdots]$ being the Gaussian bracket. The case when $\zeta$ is a quadratic irrational is important; $\zeta$ (resp\verb/./ $\Delta$) plays a similar role for the CS to that of $\tau$ (resp\verb/./ the window) for a 1D QL, and a class of CSs specified by $\zeta$ can be divided into three subclasses, I, II, and III; $\Delta$ of a CS belonging to each class is, respectively, integral, rational or irrational with respect to the module ${\bf Z}[\zeta]$. A CS has an inflation symmetry only when it belongs to subclass I or II. A type I CS is related to a type I QL such as the FL.

\begin{figure}
\begin{center}\epsfile{file=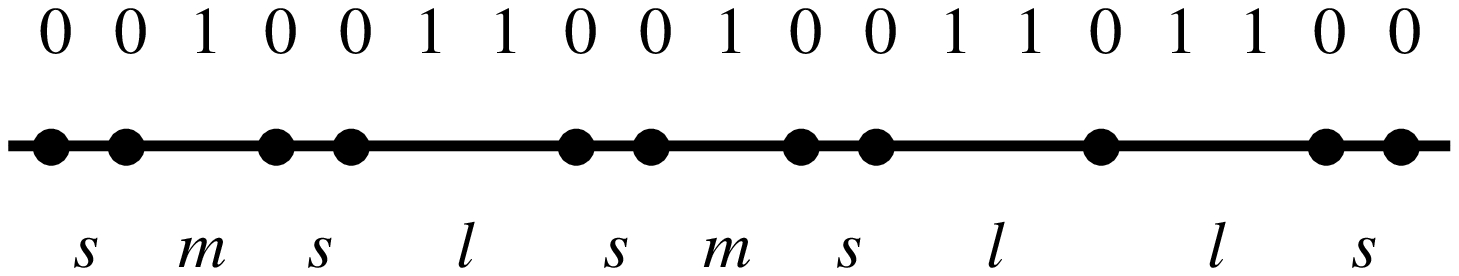,height=2.5cm}\end{center}
\caption{Geometrical relation between the circle sequence, with $\Delta = 1/2$ and $\zeta=(2-\protect\sqrt{2})/2$, and the relevant type II QL, QL(B).}
\label{fig:circleseq}
\end{figure}

It can be shown that a CS with $\Delta = 1/2$ is symmetric with respect to exchanging the two types of digits, so that two digits have an equal frequency in this CS. Moreover, if we assume further $1/4<\zeta<1/2$, each digit is isolated or paired in the CS but a triplet or higher multiplets never appear. These features can be observed, for example, in a CS shown in Fig\verb/./\ref{fig:circleseq}, for which $\Delta = 1/2$ and $\zeta=(2-\sqrt{2})/2$.

Quite recently, the authors have succeeded in revealing a general rule which relates a type II CS to a type II QL. To see the relation, we shall tentatively locate the digits of a CS consecutively on a periodic 1D lattice, and then delete all the lattice points on which digit ``1'' is located. The resulting lattice includes a quasiperiodic array of vacancies. This defective lattice is shown under a certain condition to have three types of spacings which can be denoted by the symbols $s$, $m$, and $l$ according to their lengths. This rule converts a CS into a sequence of the three symbols $s$, $m$, and $l$. We can show that the latter sequence is isomorphic to the sequence of spacing of a 1D QL. A CS can be, conversely, derived from a 1D QL. We do not, however, present here the proof of these results because a considerable space is neccesary. Anyway, four 1D QLs, QL(B), QL(C), QL(E), and
QL(F), are related by this relation to four CSs with $\Delta = 1/2$ and $\zeta=(2-\sqrt{2})/2$, $\sqrt{2}-1$, $(3-\sqrt{5})/2$, and $(\sqrt{3}-1)/2$, respectively, while QL(D) is related to a CS with $\Delta = (1+\sqrt{5})/8$ and $\zeta=(\sqrt{5}-1)/4$. The relation is
illustrated for the case of QL(B) in Fig\verb/./\ref{fig:circleseq}. A proof of the general rule relating a type II CS to a type II QL will be presented elsewhere.

Let CS(B) be the CS which is related to QL(B) by the rule above. Then CS(B) and QL(B) will belong to a common universality class with respect to the one-electron properties because each of the two quasiperiodic objects is given as a decoration of the other. A tight-binding Hamiltonian of the type of the site model is appropriate for the case of CS(B); we may assign two site energies $V_0$ and $V_1$ to the two digits of the CS. This fact has been exploited in FN to investigate one-electron properties of QL(B). We should mention in this respect that presence of gaps with unusual scaling properties was for the first time reported for CS(E) by Luck. \cite{Lu89}

In conclusion, we re-emphasize that the most remarkable feature of the one-electron properties of type II QLs is the presence of marginal critical states, which belong to a special kind of critical states with almost localized nature. The asymptotic behavior of a marginal critical state is not characterized by power-law scaling being appropriate for a conventional critical state but by a stretched exponential. As a consequence, it can be argued that a type II QL is expected to be much less conductive than a type I QL, if one assumes the same order of local modulation strength for both QLs; the difference originates from the distinct global scaling properties of these two QLs.

\section*{Acknowledgements}

The authors would like to thank Dr. Y. Ishii for a helpful comment.



\newpage

\begin{table}
\caption{A list of 1D QLs. The quadratic irrational $\tau$ characterizing the mother lattice of each QL is given in the third column, while the scaling ratio, $\sigma$, in the fourth column; $\sigma$ always takes a power of $\tau$. In the last column is given the inflation rule which specifies the self similarities of the relevant QL. Identical symbols, $s$, $m$, and $l$, do not necessarily imply the same lengths for different QLs.}
\label{table1}
\newcommand{\lw}[1]{\smash{\lower2.0ex\hbox{#1}}}
\begin{center}
\renewcommand{\arraystretch}{1.1}
\begin{tabular}{ccccl}
 & type & $\tau$ & ratio $\sigma$ & {inflation rule}\\
\hline
(A) & I & {$1+\sqrt{2}$}
 & $\tau$ & {$s' = m$, $m' = mms$}\\
(B) & II & {$1+\sqrt{2}$}
 & $\tau$ & {$s' = ms$, $m' = ls$, $l' = lls$}\\
(C) & II & {$1+\sqrt{2}$}
 & $\tau^2$ & {$s' = mms$, $m' = mmlmms$, $l' = mmlmlmms$} \\
(D) & II & $2+\sqrt{5}$
 & $\tau$ & {$s'= mss$, $m'= lslss$, $l'=lslslss$} \\
(E) & II & $\frac{1}{2}(1+\sqrt{5})$
 & $\tau^3$ & {$s'= mms$, $m'= mlms$, $l'= mlmlms$} \\
(F) & II & $2+\sqrt{3}$
 & $\tau$ & $s'= ms$, $m'= mlms$, $l'= mllms$ \\
\end{tabular}
\end{center}
\end{table}

\begin{table}
\caption{The 2D maps and the relevant parameters of the type II QLs listed in Table \protect{\ref{table1}}.}
\label{tab:dynmap}
\newcommand{\lw}[1]{\smash{\lower2.0ex\hbox{#1}}}
\begin{center}
\begin{tabular}{cclcc}
 & $\sigma$ & 2D map & $\rho$ & $\nu$ \\ \hline
(B) & $1+\sqrt{2}$ & $f'= g,\; g'= fg$ & $\frac{1}{2}(1+\sqrt{5})$ & 0.5460\\
(C) & $3+2\sqrt{2}$ & $f'= f^3 g,\; g'= f^2 g$ & $2+\sqrt{3}$ & 0.7471\\
(D) & $2+\sqrt{5}$ & $f'= fg,\; g'= f^2 g$ & $1+\sqrt{2}$ & 0.6105\\
(E) & $2+\sqrt{5}$ & $f'= fg^2,\; g'= fg$ & $1+\sqrt{2}$ & 0.6105\\
(F) & $2+\sqrt{3}$ & $f'= f^2 g,\; g'= fg$ & $\frac{1}{2}(3+\sqrt{5})$ & 0.7308
\end{tabular}
\end{center}
\end{table}


\begin{references}

\bibitem{BeMa93} E. Belin and D. Mayou, Phys. Scr. {\bf T49}, 356 (1993); D. Mayou,
in {\it Lectures on Quasicrystals}, edited by F. Hippert and D. Gratias (Les Editions
de Physique, Les Ulis, 1994).

\bibitem{Sire94} C. Sire, in {\it Lectures on Quasicrystals} (Ref. \onlinecite{BeMa93}).

\bibitem{SchBaBell98} H. Schulz-Baldes and J. Bellissard, Rev. Math. Phys. {\bf 10}, 1
(1998); J. Bellissard, Materials Science and Engineering: A {\bf 294-296}, 450 (2000).

\bibitem{RoTrMa97} S. Roche, G. Trambly de Laissardi$\grave{\rm e}$re, and D. Mayou,
J. Math. Phys. {\bf 38}, 1794 (1997).

\bibitem{Mayou} D. Mayou, Phys. Rev. Lett. {\bf 85}, 1290 (2000).

\bibitem{FuNi00}{N. Fujita and K. Niizeki, Phys. Rev. Lett. {\bf 85}, 4924 (2000).}

\bibitem{NiFu00}{K. Niizeki and N. Fujita, Report No. cond-mat/0009422.}

\bibitem{DuKa85} M. Duneau and A. Katz, Phys. Rev. Lett. {\bf 54}, 2688 (1985).

\bibitem{Ni89} See, for example, K. Niizeki, J. Phys. A: Math. Gen. {\bf 22}, 193 (1989).

\bibitem{Ba91} For the concept of MLD and its application to several 2D QLs, see M. Baake, M. Schlottmann, and P. D. Jarvis, J. Phys. A: Math. Gen. {\bf 24}, 4637 (1991); M. Baake, Report No. math-ph/9901014.

\bibitem{KKT83} M. Kohmoto, P. Kadanoff, and C. Tang, Phys. Rev. Lett. {\bf 50}, 1870 (1983).

\bibitem{KST87} M. Kohmoto, B. Sutherland, and C. Tang, Phys. Rev. B {\bf 35}, 1020 (1987).

\bibitem{HiKo92} H. Hiramoto and M. Kohmoto, Int. J. Mod. Phys. B (Singapore) {\bf 6}, 281 (1992).

\bibitem{Ha86} T. C. Halsey, M. H. Jensen, L. P. Kadanoff, I. Procaccia, and B. I. Shraiman, Phys. Rev. A {\bf 33}, 1141 (1986).

\bibitem{Ko88} M. Kohmoto, Phys. Rev. A {\bf 37}, 1345 (1988).

\bibitem{AsSt88} J. A. Ashraff and R. B. Stinchcombe, Phys. Rev. B {\bf 37}, 5723 (1988).

\bibitem{NiNo86} Q. Niu and F. Nori, Phys. Rev. Lett. {\bf 57}, 2057 (1986); Phys. Rev. B {\bf 42}, 10329 (1990).

\bibitem{KoNo90} M. Kol$\acute{\rm a}\check{\rm r}$ and F. Nori, Phys. Rev. B {\bf 42} 1062 (1990).

\bibitem{ChaKa89} S. N. Karmakar, A. Chakrabarti, and R. K. Moitra, J. Phys.: Condens. Matter {\bf 1}, 1423 (1989); A. Chakrabarti, S. N. Karmakar, and R. K. Moitra, Phys. Rev. B {\bf 39}, 9730 (1989); A. Chakrabarti and S. N. Karmakar, Phys. Rev. B {\bf 44}, 896 (1991); Phys. Lett. A {\bf 168}, 301 (1992).

\bibitem{BaLu94} D. Barache and J. M. Luck, Phys. Rev. B {\bf 49}, 15004 (1994).

\bibitem{Ko92} M. Kohmoto, J. Stat. Phys. {\bf 66}, 791 (1992).

\bibitem{Ho88} M. Holzer, Phys. Rev. B {\bf 38}, 1709 (1988).

\bibitem{FuLiZhSr97} X. Fu, Y. Liu, P. Zhou, and W. Sritrakool, Phys. Rev. B {\bf 55}, 2882 (1997).

\bibitem{BoGh93} See for example, A. Bovier and J. -M. Ghez, Commun. Math. Phys. {\bf 158}, 45 (1993); Commun. Math. Phys. {\bf 166}, 431 (1994); D. Damanik, Report No. math-ph/9912005.

\bibitem{foot} The relevant structures include circle sequences, which have a general relationship with type II 1D QLs as discussed in Sec\verb/./\ref{sec:discussions}.


\bibitem{AuGoLu88} S. Aubry, C. Godr${\rm \grave{e}}$che and J. M. Luck, J. Stat. Phys. {\bf 51}, 1033 (1988).

\bibitem{Lu89} J. M. Luck, Phys. Rev. B {\bf 39}, 5834 (1989).

\end{references}
\end{document}